\documentclass[twocolumn]{aastex631}

\usepackage{multirow}
\usepackage{cases}

\submitjournal{ApJ}

\shorttitle{Branching ratio for $\text{O}+\text{H}_3^+$ forming $\text{OH}^+ +\text{H}_2$ and $\text{H}_2\text{O}^+ +\text{H}$}
\shortauthors{Hillenbrand et al.}
\graphicspath{{./}{figures/}}

\begin{document}

\title{Branching ratio for $\text{O}+\text{H}_3^+$ forming $\text{OH}^+ +\text{H}_2$ and $\text{H}_2\text{O}^+ +\text{H}$}

\author[0000-0003-0166-2666]{Pierre-Michel Hillenbrand}
\affiliation{Columbia Astrophysics Laboratory, Columbia University, New York, NY 10027, U.S.A.}
\affiliation{Institut f\"ur Kernphysik, Goethe-Universit\"at, D-60438 Frankfurt, Germany}

\author[0000-0002-9950-4449]{Nathalie de Ruette}
\affiliation{Columbia Astrophysics Laboratory, Columbia University, New York, NY 10027, U.S.A.}
\affiliation{European Spallation Source ERIC, SE-22100 Lund, Sweden}

\author[0000-0003-3326-8823]{Xavier Urbain} 
\affiliation{Institute of Condensed Matter and Nanosciences, Universit\'e catholique de Louvain, B-1348 Louvain-la-Neuve, Belgium}

\author[0000-0002-1111-6610]{Daniel W.~Savin} 
\affiliation{Columbia Astrophysics Laboratory, Columbia University, New York, NY 10027, U.S.A.}

\begin{abstract}
The gas-phase reaction of $\mathrm{O}+\mathrm{H}_3^+$ has two exothermic product channels, $\mathrm{OH}^+ +\mathrm{H}_2$ and $\mathrm{H}_2\mathrm{O}^+ +\mathrm{H}$. In the present study, we analyze experimental data from a merged-beams measurement to derive thermal rate coefficients resolved by product channel for the temperature range from 10 to 1000~K. Published astrochemical models either ignore the second product channel or apply a temperature-independent branching ratio of 70\% vs.~30\% for the formation of $\mathrm{OH}^+ +\mathrm{H}_2$ vs.~$\mathrm{H}_2\mathrm{O}^+ +\mathrm{H}$, respectively, which originates from a single experimental data point measured at 295~K.  Our results are consistent with this data point, but show a branching ratio that varies with temperature reaching 58\% vs.~42\% at 10~K. We provide recommended rate coefficients for the two product channels for two cases, one where the initial fine-structure population of the O$(^3P_J)$ reactant is in its $J=2$ ground state and the other one where it is in thermal equilibrium. 
\end{abstract}


\section{Introduction} \label{sec:intro}

At the low temperatures typical of diffuse and dense molecular clouds ($\sim 10-100$~K), gas-phase formation of water is dominated by a sequence of exothermic ion-neutral reactions \citep{van_dishoeck_interstellar_2013}. Two parallel formation pathways have been identified: One pathway proceeds via the reaction

\begin{subnumcases}{\mathrm{O} + \mathrm{H}_3^+ \rightarrow \label{eq:OonH3+}}
        \mathrm{OH}^+ + \mathrm{H}_2 &+ 0.66~eV \label{eq:OH+}\\
        \mathrm{H}_2\mathrm{O}^+ + \mathrm{H} &+ 1.70~eV, \label{eq:H2O+}
\end{subnumcases}
which is the focus of the present paper. The quoted energies represent the exoergicities of the reactions provided by \citet{milligan_h3o_2000}.
The other pathway proceeds via cosmic-ray ionization of O to O$^+$ or endothermic charge transfer with H$^+$, produced by cosmic ray ionization of H and H$_2$, and the subsequent reaction \citep{bulut_accurate_2015,kovalenko_oh_2018}
\begin{equation}
    \mathrm{O}^+ + \mathrm{H}_2 \rightarrow \mathrm{OH}^+ + \mathrm{H}.
\end{equation}
Both pathways are followed by the reactions \citep{tran_formation_2018,kumar_low_2018}
\begin{equation}\label{eq:H3O+}
    \mathrm{OH}^+ \stackrel{\mathrm{H}_2}\longrightarrow \mathrm{H}_2\mathrm{O}^+ \stackrel{\mathrm{H}_2}\longrightarrow \mathrm{H}_3\mathrm{O}^+.
\end{equation}
Dissociative recombination (DR) of electrons with H$_3$O$^+$ then leads to the formation of H$_2$O \citep{novotny_fragmentation_2010}. Two other important ingredients for an accurate gas-phase reaction network for forming water are the DR rate coefficients leading to the destruction of OH$^+$ \citep{amitay_dissociative_1996,stroe_electron-induced_2018} and H$_2$O$^+$ \citep{rosen_recombination_2000,nkambule_theoretical_2015}.

The various routes of interstellar water chemistry, including reactions (\ref{eq:OonH3+})-(\ref{eq:H3O+}), have been reviewed by \cite{van_dishoeck_interstellar_2013,van_dishoeck_water_2021}. The branching ratio for reaction (\ref{eq:H2O+}) is particularly important for describing the ortho-to-para ratio of water in interstellar clouds \citep{faure_ortho--para_2019}. Furthermore, the observed abundances of $\mathrm{OH}^+$, $\mathrm{H}_2\mathrm{O}^+$, and $\mathrm{H}_3\mathrm{O}^+$ are used to infer the cosmic-ray ionization rate \citep{hollenbach_chemistry_2012, indriolo_herschel_2015, neufeld_cosmic-ray_2017, indriolo_constraints_2018}. At temperatures of $\sim250$~K, endothermic neutral-neutral reactions such as $\mathrm{O}+\mathrm{H}_2\rightarrow \mathrm{OH}+\mathrm{H}$ become accessible and dominate the gas-phase water formation process \citep{van_dishoeck_water_2021}. This sets an upper limit for the temperature range of interest for reactions (\ref{eq:OonH3+})-(\ref{eq:H3O+}).

In the present paper, we investigate the temperature-dependent thermal rate coefficients and branching ratios for reactions (\ref{eq:OH+}) and (\ref{eq:H2O+}). We derive our results from the merged-beams experiment performed by \citet{de_ruette_merged-beams_2016}. In their work, they derived the thermal rate coefficient for the sum over both product channels, but not those resolved by product channel. We provide the relevant details of the experimental data set in Sec.~\ref{sec:exp}. In Sec.~\ref{sec:thermal} we present our results and compare them to the data available in the literature. Lastly, in Sec.~\ref{sec:implications} we discuss possible implications of our results on astrochemical models.

\section{Experimental data}\label{sec:exp}

In the merged-beams experiment of \citet{de_ruette_merged-beams_2016}, a beam of atomic oxygen was overlapped with a beam of H$_3^+$ at a well-defined relative collision energy. The product ions, either OH$^+$ or H$_2$O$^+$, were counted, normalized to the intensities of the two parent beams and the interaction volume, and corrected for transmittance and detection efficiencies. The particle densities in the interaction volume were low enough to exclude the formation of H$_2$O$^+$ by sequential collisions.

From the data, absolute cross sections as a function of the relative collision energy, $E_\mathrm{r}$, were derived for each product channel individually. The range of $E_\mathrm{r}$ was 3.5~meV to 15.5~eV and 3.5~meV to 130~meV for the OH$^+$ and H$_2$O$^+$ products, respectively. At values of $E_\mathrm{r}$ higher than the ones investigated, the individual cross sections were smaller than the detection sensitivity of the experiment. 

The data for each product channel was parameterized by the fit function
\begin{equation}\label{eq:cs}
    \sigma_x (E_\mathrm{r}) =\frac{a_0  + a_{1/2} E_\mathrm{r}^{1/2}}{E_\mathrm{r}^{2/3}+b_1 E_\mathrm{r} + b_2 E_\mathrm{r}^2 + b_4 E_\mathrm{r}^4}.
\end{equation}
Here, $x=\{\mathrm{OH}^+,~\mathrm{H}_2\mathrm{O}^+\}$ refers to reactions (\ref{eq:OH+}) and (\ref{eq:H2O+}), respectively, $\sigma_x$ is in units of cm$^2$, and $E_\mathrm{r}$ is in eV. The $\sim E_\mathrm{r}^{-2/3}$ scaling in the limit of $E_\mathrm{r}\rightarrow 0$ reflects the effect of the long-range charge-quadrupole interaction between the $\mathrm{H}_3^+$ ion and the oxygen atom, as was discussed by \citet{klippenstein_temperature_2010}. The fit parameters are given in Tab.~\ref{tab:cs}. The total uncertainty of the cross sections was given by the fitting uncertainty and the relative systematic uncertainty of 13\%, both summed in quadrature. 

Considering the astrochemical motivation of this study and the range of measured collision energies, we derived translational rate coefficients, $k_x^\mathrm{tr}(T)$, in the temperature range of $T=10-1000$~K by convoluting each cross section times the relative collision velocity, $\sigma_x (E_\mathrm{r})\times v_\mathrm{r}$, with a Maxwell-Boltzmann distribution. The uncertainties of the rate coefficients were propagated from the total uncertainties of the cross sections. Furthermore, we define the branching ratios, $f_x(T)$, as the relative yields,
\begin{subequations}\label{eq:branching}
\begin{eqnarray}
    f_{\mathrm{OH}^+} &=& \frac{k^\mathrm{tr}_{\mathrm{OH}^+}}{k^\mathrm{tr}_{\mathrm{OH}^+}+k^\mathrm{tr}_{\mathrm{H}_2\mathrm{O}^+}}, \\
    f_{\mathrm{H}_2\mathrm{O}^+} &=& \frac{k^\mathrm{tr}_{\mathrm{H}_2\mathrm{O}^+}}{k^\mathrm{tr}_{\mathrm{OH}^+}+k^\mathrm{tr}_{\mathrm{H}_2\mathrm{O}^+}}. 
\end{eqnarray}    
\end{subequations}
The uncertainty of $f_x(T)$ is solely affected by statistical fluctuations, while the systematic uncertainty cancels out. In the considered temperature range, the relative uncertainties of $f_{\mathrm{OH}^+}$ and $f_{\mathrm{H}_2\mathrm{O}^+}$ range from 3 to 5\% and 5 to 12\%, respectively, while the absolute magnitude of the uncertainty is identical for both branching ratios.

To understand the relation between $k_x^\mathrm{tr}(T)$ and the thermal rate coefficient, $k_x(T)$,  we will discuss the role of the internal excitation of the two reactants in Secs.~\ref{sec:H3+} and \ref{sec:O}, respectively. 

\begin{deluxetable}{ccc}
	\tablecaption{Parameters for $\sigma_x (E_\mathrm{r})$ defined in Eq.~(\ref{eq:cs}).\label{tab:cs}}
	\tablehead{
	\colhead{Parameter} & \multicolumn2c{Product channel $x$}\\
	\cline{2-3}
	 & \colhead{$\mathrm{OH}^+ + \mathrm{H}_2$\tablenotemark{$\ast$}} & \colhead{$\mathrm{H}_2\mathrm{O}^+ + \mathrm{H}$}}
	\startdata
	$a_0$ & $1.1880\times10^{-16}$ & $9.8531\times10^{–17}$ \\
	$a_{1/2}$ & $5.1976\times10^{–16}$ & - \\
	$b_1$ & - & $-4.0668\times10^{–2}$ \\
	$b_2$ & $7.3434\times10^{–2}$ & $-4.1891$ \\
	$b_4$ & $8.3488\times10^{-4}$ & $5.1780\times10^2$
	\enddata
	\tablenotetext{\ast}{The values for this reaction were misprinted in \citet{de_ruette_merged-beams_2016}.}
\end{deluxetable}

\subsection{The $\mathrm{H}_3^+$ reactant}\label{sec:H3+}
 
The merged-beams experiment of \citet{de_ruette_merged-beams_2016} was performed with an internally excited H$_3^+$ beam. Most of the possible influences of this internal excitation on the measured merged-beams rate coefficient were discussed in their publication. In the experiment, the internal temperature of the H$_3^+$ ions was estimated to be $T_\mathrm{int}\approx2500-3000$~K. In \citet{hillenbrand_experimental_2019}, we performed a more detailed analysis of the H$_3^+$ internal temperature and its dependency on the H$_2$ pressure in the ion source. From these results we confirmed that the H$_2$ pressure applied in the measurements of \citet{de_ruette_merged-beams_2016} corresponds consistently to the previously estimated internal temperature. Here, we discuss an additional aspect of the H$_3^+$ internal temperature that was not considered in \citet{de_ruette_merged-beams_2016}.  

For comparison, the exothermic reaction of $\mathrm{C}+\mathrm{H}_2^+\rightarrow \mathrm{CH}^+ + \mathrm{H}$ was studied in \citet{hillenbrand_experimental_2020}. In that reaction, a low-lying endoergic reaction channel forming $\mathrm{C}^+ + \mathrm{H} + \mathrm{H}$ is energetically accessible at zero translational energy, provided that the internal excitation of the H$_2^+$ reactant is sufficiently high. We found that this considerably reduced the measured merged-beams rate coefficient for all translational energies investigated. 

For the reaction studied here, the lowest endoergic channel is \citep{de_ruette_merged-beams_2016}
\begin{equation}
    \mathrm{O}+\mathrm{H}_3^+ \rightarrow \mathrm{OH}+\mathrm{H}_2^+ - 1.77~\mathrm{eV}.
\end{equation}
We can estimate the fraction of $\mathrm{H}_3^+$  ions, whose internal energy is above this threshold,  $E_\mathrm{int} > E_\mathrm{th}=1.77$~eV, for a given internal temperature of $T_\mathrm{int}$ by integrating the Boltzmann distribution over the internal energy,
\begin{eqnarray}
    f_{E_\mathrm{int}>E_\mathrm{th}} & = &
    1-\frac{1}{\langle E_\mathrm{int} \rangle}  \int_0^{E_\mathrm{th}} \exp\left(-\frac{E_\mathrm{int}}{\langle E_\mathrm{int} \rangle}\right) d E_\mathrm{int} \nonumber \\
    & = & \exp \left(-\frac{E_\mathrm{th}}{\langle E_\mathrm{int} \rangle}\right).
\end{eqnarray}
Based on the partition function provided by \citet{kylanpaa_first-principles_2011}, $T_\mathrm{int}=2500$ and 3000~K correspond to mean internal energies of $\langle E_\mathrm{int} \rangle=0.64$ and 0.86~eV, where 6\% to 13\% of the $\mathrm{H}_3^+$ ions have an internal energy above $E_\mathrm{th}$, respectively. A part of this fraction can potentially form $\mathrm{OH}+\mathrm{H}_2^+$, even at the lowest measured translational energies. As a worst-case scenario, we can assume that all reactions of H$_3^+$ ions with $E_\mathrm{int} > E_\mathrm{th}$ do not contribute to reactions (\ref{eq:OH+}) or (\ref{eq:H2O+}).  This would mean that the cross sections and rate coefficients reported by \citet{de_ruette_merged-beams_2016} need to be scaled up by a factor 1.06 to 1.15.  This scaling lies well within the total experimental uncertainty limits.  Moreover, it affects both channels in the same direction and therefore does not affect the branching ratios.

\subsection{The $\mathrm{O}\left(^3P_J\right)$ reactant}\label{sec:O}

In \citet{de_ruette_merged-beams_2016}, the atomic oxygen beam was generated through laser detachment of an oxygen anion beam \citep{oconnor_generation_2015}. The resulting oxygen atoms were in the $^3P_J$ ground term and the fine structure levels were populated according to their multiplicity, with fractional populations being 5/9, 3/9, and 1/9 for $J=2$, 1, and 0, respectively. This statistical population was verified at the few percent level in the experiment of  \citet{genevriez_absolute_2018}.

For O$(^3P_J)$, $J=2$ represents the ground level, while the $J=1$ and the $J=0$ levels are excited by $E_1=19.6$~meV and $E_0=28.1$~meV, respectively \citep{lique_fine-structure_2018}. The three fine-structure levels comprise a total of nine magnetic sublevels. As first pointed out by \citet{gentry_long-range_1977}, only the three sublevels corresponding to $J=2$ with $M_J=\pm1$ or $M_J=0$ correlate to the reactive $^3\Sigma$ potential energy surface (PES). The other six sublevels all correlate with the nonreactive $^3\Pi$ PES \citep{gentry_long-range_1977,bettens_interpolated_1999,klippenstein_temperature_2010}. Since both product channels only proceed on the $^3\Sigma$ PES, the branching ratio of the product channels is independent of the initial fine-structure population. This aspect was overlooked by \citet{de_ruette_merged-beams_2016} and prevented them from deriving thermal rate coefficients for  the individual product channels.

Two astrochemical scenarios for the fine-structure level population may be of interest: Either only the $J=2$ ground level is populated or the fine-structure level populations are in thermal equilibrium.  Which of the two cases is applicable actually depends on the particle density. According to \citet{lique_fine-structure_2018}, atomic oxygen is thermalized only at H or H$_2$ densities on the order of $10^6-10^7~\textrm{cm}^{-3}$. 

The conversion of the measured translational rate coefficients into thermal rate coefficients assuming a statistical population of the fine-structure levels in the merged-beams experiment and a thermal population in the astrochemical environment is \citep{de_ruette_merged-beams_2016}
\begin{eqnarray}\label{eq:fs}
    k_x(T)~& = &~k_x^\mathrm{tr}(T) \\ 
    & \times & \frac{9}{5+3\exp\left(-E_1/k_\mathrm{B}T\right)+\exp\left(-E_0/k_\mathrm{B}T\right)}.
    \nonumber
\end{eqnarray}
Here, $k_\mathrm{B}$ is the Boltzmann constant. For the scenario of a pure O$(^3P_2)$ reactant, the relation reduces to $k_x(T)=k_x^\mathrm{tr}(T) \times 9/5$. As discussed above, the rate coefficients for O$(^3P_1)$ and O$(^3P_0)$ reactants are zero for both scenarios.

\section{Thermal rate coefficient}\label{sec:thermal}

\subsection{Summed over both product channels}

The thermal rate coefficient summed over both product channels, assuming a thermal population of the $\mathrm{O}\left(^3P_J\right)$ fine-structure levels, has been discussed in detail by \citet{de_ruette_merged-beams_2016}. The temperature dependency of the thermal rate coefficient is mainly formed by the effect of the long-range charge-quadrupole interaction [see Eq.~(\ref{eq:cs})], and the effect of the fine-structure population [see Eq.~(\ref{eq:fs})]. The results were compared to two published experimental data points, both measured at room temperature: \citet{fehsenfeld_ion_1976} used a flowing afterglow technique to measure the thermal rate coefficient summed over both product channels, and \citet{milligan_h3o_2000} applied a combined flowing afterglow/selected flow tube technique to measure thermal rate coefficients resolved by product channel. Both methods are such that the translational temperature and the internal temperature of both reactants were in thermal equilibrium. The results of \citet{de_ruette_merged-beams_2016} were in good agreement with the measurement of \citet{fehsenfeld_ion_1976} and in reasonable agreement with the measurements of \citet{milligan_h3o_2000}. Furthermore, the theoretical rate coefficients of \citet{bettens_interpolated_1999} and \citet{klippenstein_temperature_2010} had the same temperature dependency as the result of \citet{de_ruette_merged-beams_2016}, but were overall a factor of about 1.5 higher. Following the discussion of Sec.~\ref{sec:H3+} it is unlikely that such a discrepancy was caused by the internal excitation of the H$_3^+$ beam in the experiment.

\begin{figure}
\includegraphics[width=1.0\columnwidth]{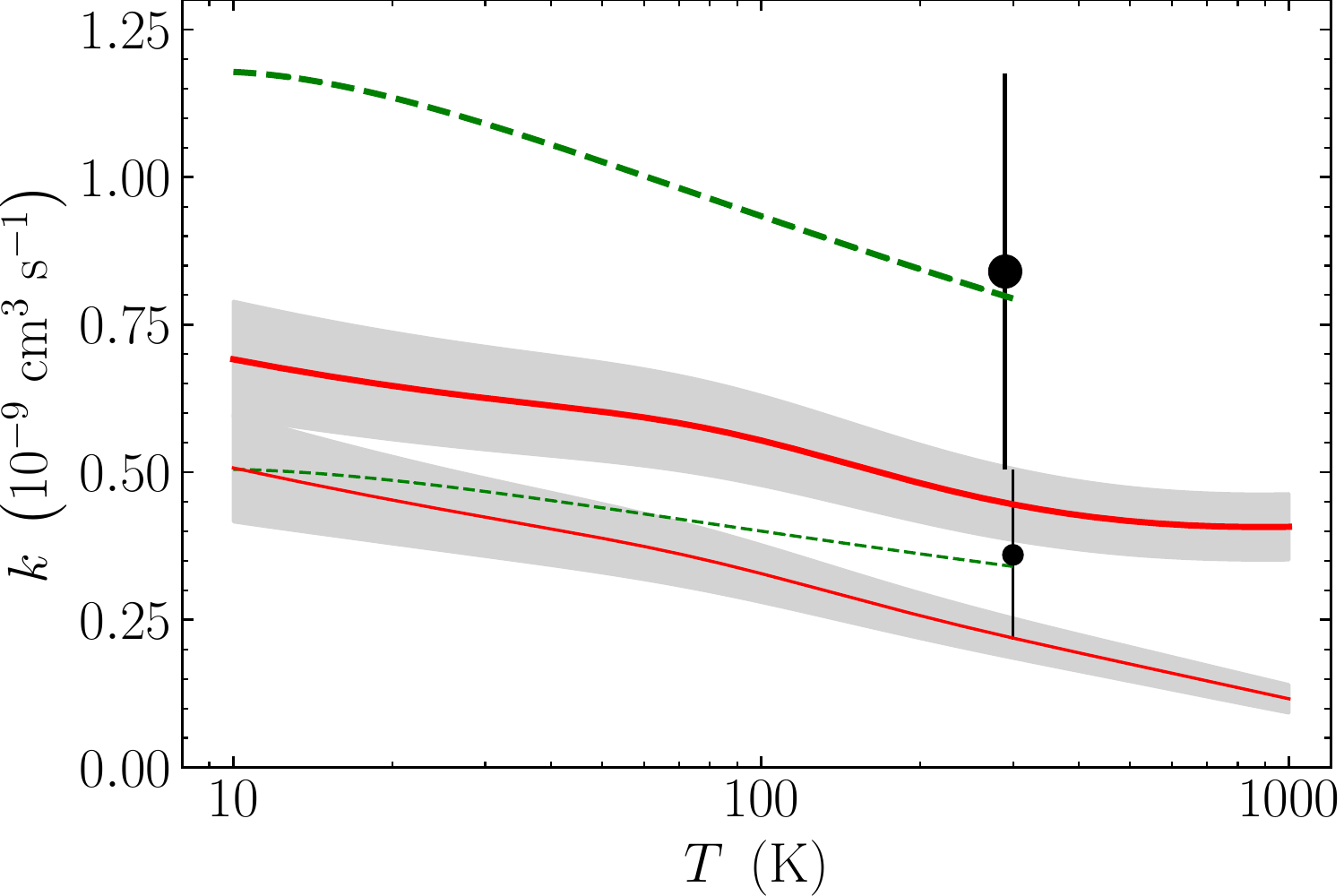}
\caption{Thermal rate coefficient for the $\mathrm{OH}^+ +\mathrm{H}_2$ channel (upper bold data) and the $\mathrm{H}_2\mathrm{O}^+ +\mathrm{H}$ channel (lower fine data). Shown are: our results by the red solid lines, with the gray area indicating the total experimental uncertainty at one standard deviation; the results of \citet{milligan_h3o_2000} by the circular data points (shifted slightly in temperature for clarity); and the data implemented in the astrochemical models listed in Sec.~\ref{sec:channel} by the green dashed lines.
\label{fig:channels}}
\end{figure}

\subsection{Resolved by product channel}\label{sec:channel}

The thermal rate coefficient resolved by product channel, derived from the data of \citet{de_ruette_merged-beams_2016}, are shown in Fig.~\ref{fig:channels}. Considering the mutual uncertainties, they agree reasonably well with the measurement of \citet{milligan_h3o_2000} at 295~K. Astrochemical models such as UMIST \citep{mcelroy_umist_2013}, KIDA \citep{wakelam_2014_2015}, and UGAN \citep{hily-blant_modelling_2018} implemented channel resolved rate coefficients using the total reaction rate coefficient of \cite{klippenstein_temperature_2010}, the branching ratio from \citet{milligan_h3o_2000}, and assumed the ratio to be temperature-independent. Our results are generally lower than the rates implemented in the models. Of particular note is the factor 1.7 difference at 10~K for the $\mathrm{OH}^+ + \mathrm{H}_2$ channel. 

\begin{figure}
\includegraphics[width=1.0\columnwidth]{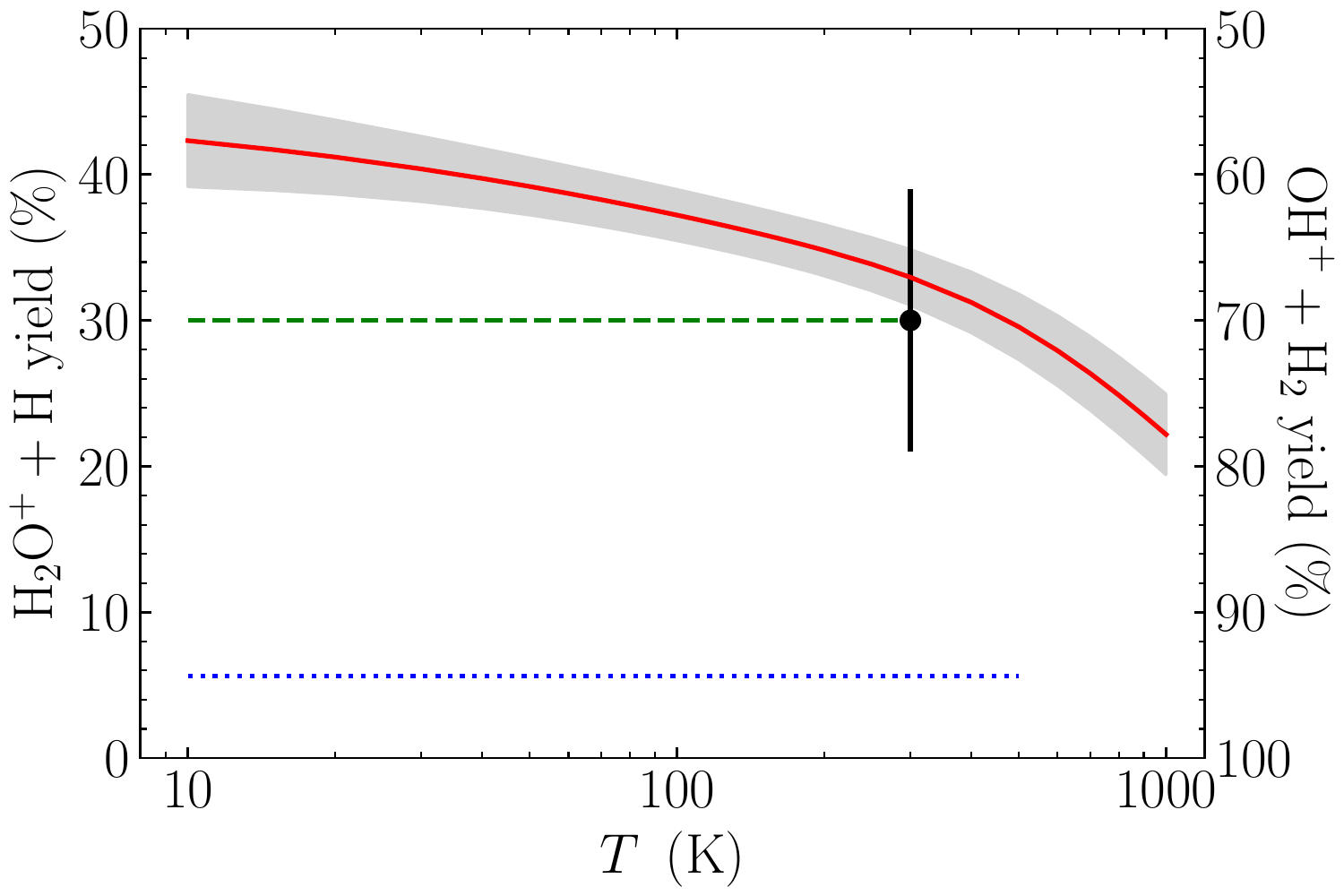}
\caption{Branching ratio expressed as relative yield for the two product channels. Shown are: our results by the red solid line, with the gray area indicating the total experimental uncertainty at one standard deviation; the results of \citet{milligan_h3o_2000} by the circular data point; the results of \citet{bettens_interpolated_1999} by the blue dotted line; and the ratio implemented in the astrochemical models listed in Sec.~\ref{sec:channel} by the green dashed line.
\label{fig:branching}}
\end{figure}

\subsection{Branching ratio}\label{sec:branching}

The branching ratio defined in Eq.~(\ref{eq:branching}) is plotted in Fig.~\ref{fig:branching}. For the $\mathrm{H}_2\mathrm{O}^+ + \mathrm{H}$ product channel the branching ratio derived from our results ranges from $(42\pm3)\%$ at 10~K down to $(22\pm3)\%$ at 1000~K. At 295~K, our result of $(33\pm2)\%$ is consistent with the value of $(30\pm9)\%$ measured by \citet{milligan_h3o_2000}.  Up to now, those astrochemical models that considered the branching ratio assumed it to be 30\% independent of the temperature, based on the measurement of \citet{milligan_h3o_2000}. 

The branching ratio calculated by \citet{bettens_interpolated_1999} was $(5.6\pm 1.6)\%$ for the $\mathrm{H}_2\mathrm{O}^+ + \mathrm{H}$ product channel, and thus does not agree at all with either set of experimental results. For the analog reaction of $\mathrm{C}+\mathrm{H}_3^+$, we found a similar discrepancy between our experimental results in \citet{oconnor_reaction_2015} and the theoretical results of \citet{bettens_potential_1998,bettens_erratum_2001}. These findings may suggest a shortcoming of the classical trajectory simulations and the PES used. The good agreement for the branching ratio for $\mathrm{O}+\mathrm{H}_3^+$ from the experimental results of \citet{milligan_h3o_2000} and our results in \citet{de_ruette_merged-beams_2016}, which used two different experimental methods with differing systematics, and for $\mathrm{C}+\mathrm{H}_3^+$ from \citet{savic_reactions_2005} and ours in \citet{oconnor_reaction_2015}, which also used two different experimental methods with differing systematics, suggest that the discrepancy is not an experimental issue.

\begin{deluxetable}{clcrr}
	\tablecaption{\label{tab:fit}Recommended thermal rate coefficients for $\mathrm{O}+\mathrm{H}_3^+$, given as parameters of Eq.~(\ref{eq:arr}) for $T=10-1000$~K.}
	\tablehead{
		\colhead{O$(^3P_J)$} & \colhead{product} & \colhead{$\alpha$} & \colhead{$\beta$} & \colhead{$\gamma$}\vspace{-0.2cm} \\ 
		\colhead{population} & \colhead{channel} & \colhead{($10^{-9}$~cm$^3$~s$^{-1}$)} & & \colhead{(K)}}
	\startdata
	\multirow{2}{*}{$J=2$} & $\mathrm{OH}^+ +\mathrm{H}_2$ & 0.612 & 0.05 & -3.08 \\
	& $\mathrm{H}_2\mathrm{O}^+ +\mathrm{H}$ & 0.271 & -0.21 & 0.57 \\
	\hline
	\multirow{2}{*}{$J$ thermal} & $\mathrm{OH}^+ +\mathrm{H}_2$ & 0.465 & -0.14 & 0.67 \\
    & $\mathrm{H}_2\mathrm{O}^+ +\mathrm{H}$ & 0.208 & -0.40 & 4.86
	\enddata
\end{deluxetable}

\subsection{Recommended rate coefficients}

We used the Arrhenius-Kooij formula,
\begin{equation}\label{eq:arr}
    k(T)=\alpha \left( \frac{T}{300~\mathrm{K}}\right)^\beta \exp\left(-\frac{\gamma}{T}\right),
\end{equation}
to parameterize our results for the two product channels in the temperature range of 10 to 1000~K for implementation into astrochemical models. The fit parameters, $\alpha$, $\beta$, and $\gamma$, are shown in Tab.~\ref{tab:fit} for the two cases of the O$(^3P_J)$ level population discussed in Sec.~\ref{sec:O} and the two product channels. The fits describe our results to better than 5\% for reaction (\ref{eq:OH+}) and to better than 10\% for reaction (\ref{eq:H2O+}).

Resolved by the nuclear-spin states, reactions (\ref{eq:OH+}) and (\ref{eq:H2O+}) read 
\begin{subnumcases}{\mathrm{O} + \textrm{p-H}_3^+ \rightarrow \label{eq:pOonH3+}}
        \mathrm{OH}^+ + \textrm{p-H}_2 & [1/2] \label{eq:pH2}\\
        \mathrm{OH}^+ + \textrm{o-H}_2 & [1/2] \label{eq:oH2}\\
        \textrm{p-H}_2\mathrm{O}^+ + \mathrm{H} & [1/2]\label{eq:pH2O}\\
        \textrm{o-H}_2\mathrm{O}^+ + \mathrm{H} & [1/2]\label{eq:oH2O}
\end{subnumcases}
and
\begin{subnumcases}{\mathrm{O} + \textrm{o-H}_3^+ \rightarrow \label{eq:oOonH3+}}
        \mathrm{OH}^+ + \textrm{o-H}_2 & [1] \label{eq:oOH+}\\
        \textrm{o-H}_2\mathrm{O}^+ + \mathrm{H} & [1], \label{eq:oH2O+}
\end{subnumcases}
where p- signifies para and o- signifies ortho. In reactions (\ref{eq:pOonH3+}) and (\ref{eq:oOonH3+}), the numbers given in squared brackets represent the statistical factors resulting from the nuclear-spin selection rules \citep{oka_nuclear_2004}. In nuclear-spin state resolved astrochemical networks, the rate coefficients are typically scaled by multiplying the parameters $\alpha$ with the statistical factor while the parameters $\beta$ and $\gamma$ are considered to be independent of the spin states \citep{majumdar_chemistry_2017,hily-blant_modelling_2018}. This approach is justified for reactions where the exoergicity does not differ significantly between different spin-resolved product channels. The same approach can be applied to our results given in Tab.~\ref{tab:fit} and the resulting nuclear-spin-resolved rate coefficients can be implemented into models studying the ortho-to-para ratio of gas-phase formed water \citep{faure_ortho--para_2019}.

\section{Astrochemical implications}\label{sec:implications}

In \citet{de_ruette_merged-beams_2016}, the derived thermal rate coefficient summed over both product channels was implemented into the KIDA model \citep{wakelam_2014_2015}, and potential implications were studied for the predicted abundances of $\mathrm{H}_3\mathrm{O}^+$, $\mathrm{H}_2\mathrm{O}$, and other molecules downstream the reaction chain. Here, we briefly  discuss the relevance of the $\mathrm{O}+\mathrm{H}_3^+$ branching ratio along the following simplified reaction chain:
\begin{equation}
\mathrm{O} \stackrel{\mathrm{H}_3^+}\longrightarrow \left\{
\begin{array}{l}
    \mathrm{OH}^++ \mathrm{H}_2\left\{\begin{array}{l}
        \stackrel{\mathrm{H}_2}\longrightarrow \mathrm{H}_2\mathrm{O}^+ + \mathrm{H} + \mathrm{H}_2\\
        \stackrel{\mathrm{e}^-}\longrightarrow \mathrm{O}+\mathrm{H} + \mathrm{H}_2
    \end{array}\right. \\
    \mathrm{H}_2\mathrm{O}^+ + \mathrm{H}. 
\end{array}\right.
\end{equation}

For typical astrochemical environments, the $\mathrm{H}_2$ density is high enough such that the $\mathrm{H}_2\mathrm{O}^+$ formation through the intermediate step of $\mathrm{OH}^+$ formation proceeds with an effective rate coefficient comparable to the direct formation pathway. In this case, including the branching ratio decreases the predicted $\mathrm{OH}^+$ abundance, but not the $\mathrm{H}_2\mathrm{O}^+$ abundance nor subsequent products along the reaction chain, such as 
$\mathrm{H}_3\mathrm{O}^+$ and, prominently, $\mathrm{H}_2\mathrm{O}$. However, in those scenarios where the destruction of $\mathrm{OH}^+$ by DR plays a significant role, direct formation of $\mathrm{H}_2\mathrm{O}^+$ presents a bypass with respect to the $\mathrm{OH}^+$ destruction by DR, and thus may enhance the abundance of $\mathrm{H}_2\mathrm{O}^+$ and its subsequent reaction products. 

\section{Summary}
Based on the experimental data of \citet{de_ruette_merged-beams_2016}, we have derived temperature-dependent thermal rate coefficients for reactions (\ref{eq:OH+}) and (\ref{eq:H2O+}). The experimental data were measured with internally excited H$_3^+$ reactants, but we find that this has no significant effect on the derived rate coefficients.

Up to now, the general understanding for the branching ratio of these reactions relied solely on the fixed temperature measurement of \citet{milligan_h3o_2000} at 295~K. At that temperature, we found a branching ratio that is consistent with the one implemented in commonly used astrochemical models, but we find an enhanced yield for the $\mathrm{H}_2\mathrm{O}^+ + \mathrm{H}$ product channel towards lower temperatures. Our results underline the importance of including both product channels into reliable models.

\begin{acknowledgments}
This research was supported, in part, by the NSF Division of Astronomical Sciences Astronomy and Astrophysics Grants program under AST-2002461. X.~U.~is a Senior Research Associate of the Fonds de la Recherche Scientifique-FNRS.
\end{acknowledgments}

\bibliography{OonH3+}{}
\bibliographystyle{aasjournal}

\end{document}